# Opposite pressure effects on magnetic phase transitions in $NiBr_2$

Parvez Ahmed Qureshi[1]*, Krishna Kumar Pokhrel[1], Jiří Prchal[1], Subhasmita Ray[1], Sergiu Arapan[2], Karel Carva[1], Vladimír Sechovský[1], and Jiří Pospíšil[1]

*[1]Charles University, Faculty of Mathematics and Physics, Department of Condensed Matter Physics, Ke Karlovu 5, 4 121 16 Prague 2, Czech Republic 5*

*[2]IT4Innovations, VŠB Technical University of Ostrava, 17. listopadu 3172/15, Ostrava 70800, Czech Republic*

## ABSTRACT

$NiI_2$ and $NiBr_2$ are archetypal van der Waals (vdW) triangular-lattice multiferroics that host incommensurate helimagnetic order at the lowest temperatures and undergo a transition to collinear antiferromagnetic order upon heating. Focusing on $NiBr_2$, we reveal that both antiferromagnetic phases exhibit a pronounced sensitivity to hydrostatic pressure. The Néel temperature of the collinear phase increases steeply at ~20 K/GPa, reaching ~100 K at 3 GPa without any indication of saturation, whereas the helimagnetic phase is completely suppressed only above ~0.8 GPa. This behavior contrasts sharply with $NiI_2$, in which both helical and collinear phases are strengthened until a moderate pressure of ~6 GPa, above which the helical phase instantly disappears. Ab initio calculations identify the second-nearest interlayer exchange interaction ($j_2'$) as the primary driver stabilizing the collinear AFM phase in $NiBr_2$. In addition, the in-plane exchange ratio renders the helical order in $NiBr_2$ considerably more fragile, enabling its suppression under relatively small pressures. These results underscore the dominant role of interlayer interactions in governing the distinct pressure responses of the magnetic phases in $NiBr_2$ and $NiI_2$.

## INTRODUCTION

Since the discovery of graphene in 2004[1], research on 2D vdW materials has become a subject of tremendous interest and progress[2]. Materials exhibiting multiferroicity in the monolayer limit have been studied extensively from the viewpoint of fundamental physics and spintronics device applications.[3]. The discovery of multiferroicity in a $NiI_2$ monolayer[4] boosted efforts to find further vdW materials exhibiting stable ferroelectric[5, 6] or, ideally, multiferroic properties[7, 8]. Several room-temperature vdW magnets have recently been discovered[9-11], but intrinsic room-temperature vdW multiferroics remain elusive.

Pressure provides a powerful tool for tuning magnetic interactions in condensed matter systems by altering interatomic distances, orbital overlap, and exchange coupling pathways without introducing chemical disorder. In vdW AFM materials, the pressure response becomes particularly intriguing due to the coexistence of strong in-plane covalent bonding and weak interlayer van der Waals coupling[12]. vdW AFM materials generally show significant increases in Néel temperature ($T_N$) under applied pressure, with pressure coefficients $dT_N/dP$ typically ranging from ~5 to ~15 K/GPa depending on the material's electronic structure and magnetic exchange hierarchy[13-15]. An enormous increase in the transition temperature to a multiferroic phase in $NiI_2$ has been recently reported by applying external pressure [15]. Two distinct magnetic phase transitions are observed in $NiI_2$ upon cooling. First, a colinear AFM state is established below Neel temperature $T_{N1}$ = 76 K, transforming at $T_{N2}$ = 59 K into an incommensurate helimagnetic (HM) phase accompanied by multiferroic properties[16, 17]. Applying high hydrostatic pressure lift $T_{N1}$ to ~200 K at 9 GPa, $T_{N2}$ increases to 120 K at a pressure of 6.5 GPa and suddenly disappears above this value[15].

$NiBr_2$ is isostructural with $NiI_2$. Both crystallize in the trigonal structure of the $CdCl_2$ type (space group R-3*m*)[18] with a triangular Ni sublattice promoting magnetic frustration[19] and exhibit similar magnetic and multiferroic properties. $NiBr_2$ orders in a collinear AFM in-plane structure below $T_{N1}$ = 44 K and transforms with cooling to an incommensurate HM phase at $T_{N2}$ = 23 K[20]. The magnetic phase transition at $T_{N2}$ is accompanied by inversion symmetry breaking, leading to ferroelectric polarization and multiferroicity [21]. The $T_{N2}$ transition was reported to be of first-order type[22].

The previously reported response of magnetic phase transitions in $NiBr_2$ to applied hydrostatic pressure is entirely different from the observations for $NiI_2$. $T_{N2}$ quickly decreases with increasing pressure, and the HM phase has been predicted to vanish at pressures above 1 GPa based on an extrapolation from measurements up to 0.5 GPa[23, 24]. Despite the experimental observations, the calculated intra- and interlayer exchange parameters predicted only the HM order across the entire studied pressure range of 0–15 GPa, with no indication of a transition to the AFM state [25]. These contradictory results on $NiBr_2$ motivated us to experimentally investigate the pressure effect on the magnetic phase transitions using AC magnetic susceptibility measurements and to revisit the theoretical electronic structure calculations with a focus on these effects, including extrapolation to $NiI_2$ behavior.

Applied pressure is generally expected to increase in interlayer exchange interactions in vdW materials, consistent with the observed increase of $T_{N1}$. Without a deeper understanding, it may be puzzling why the evolution of $T_{N2}$ with pressure exhibits the opposite trend. Basic rules for the preference of HM or AFM order can be obtained from the mean-field model applied to the Heisenberg Hamiltonian for the layered triangular lattice[26]. To apply these rules to the studied problem, one needs to know which terms in the Hamiltonian are relevant and how they evolve with pressure.

For $NiI_2$, mapping of first principles calculations to the Heisenberg spin model together with Monte Carlo simulations has been used to explain the pressure-induced magnetic transformations from helimagnetic to collinear antiferromagnetic present near 7 GPa[15]. First-principles calculations in $NiI_2$ have also revealed a sizable contribution of Kitaev and biquadratic interactions[27]. However, the magnitude of these terms depends strongly on relativistic effects in ligands, and these effects are expected to be much weaker in the case of Br. Finite temperature magnetism in $NiBr_2$ under pressure has been studied by first-principle calculations, together with Monte Carlo simulations, recently[25], but an answer to the critical question of the loss of helimagnetic order has not been provided.

This work presents the first comprehensive pressure-dependent study of $NiBr_2$ up to 3 GPa, revealing opposite trends in the magnetic transition temperatures. We observe a rapid increase in the AFM Néel temperature ($T_{N1}$) reaching ~100 K at 3 GPa, while the helimagnetic phase ($T_{N2}$) is simultaneously suppressed beyond 0.8 GPa. We also discuss how subtle differences in the evolution of exchange parameters with pressure can change the magnetic response in $NiBr_2$ compared to isostructural $NiI_2$, where both phases are enhanced with pressure[15]. This opposite response is unique among vdW magnets and is explained by ab initio calculations.

## RESULTS AND DISCUSSION

The temperature dependence of the specific heat of $NiBr_2$ reveals a small peak-like anomaly at $T_{N1}$ = 44 K, marking the onset of the transition between the paramagnetic (PM) and collinear AFM phase. However, the second magnetic transition at $T_{N2}$ ~ 23 K, associated with the onset of the HM phase with cooling, is not resolved in the specific heat data (Fig. 1), similar to the report of other authors [28]. This result is striking because we deal with a first-order transition[22] and $NiI_2$ exhibits significant anomalies both at $T_{N1}$ and $T_{N2}$ in specific heat data[16]. The negligible thermodynamic response at $T_{N2}$ makes the specific heat an unsuitable quantity for the pressure study of $NiBr_2$.

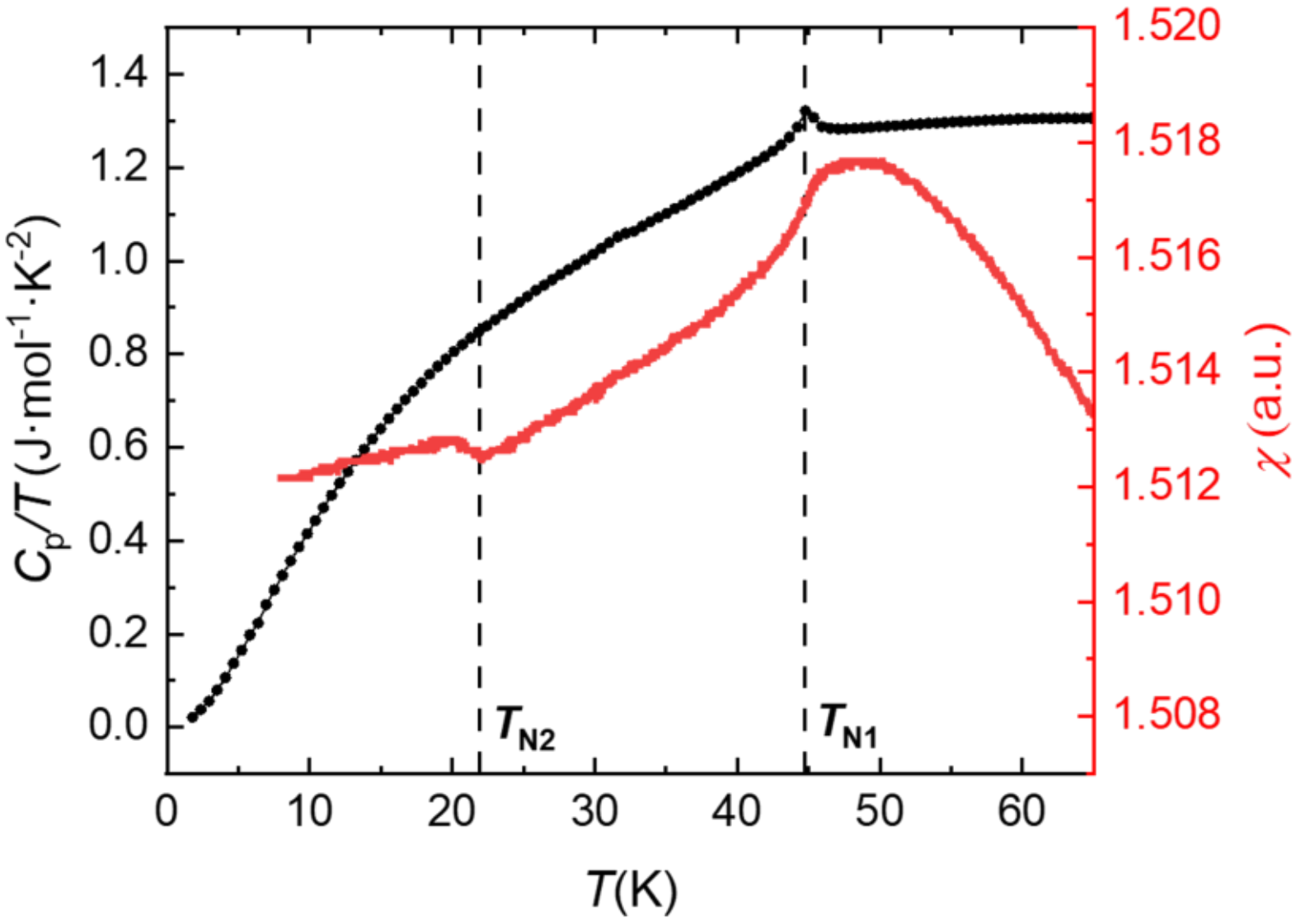

**Fig. 1| Magnetic transition in $NiBr_2$.** Temperature dependence of zero magnetic field specific heat ($C_p$/T) and temperature dependence of AC magnetic susceptibility (χ) measured with the AC magnetic field (0.3 mT) applied perpendicular to the *c*-axis of $NiBr_2$.

In contrast, AC magnetic susceptibility distinctly captures both magnetic transitions in $NiBr_2$, as reported before[20]. $T_{N1}$ was identified as the maximum slope in the data, with a tendency to sharpen with increasing pressure. On the other hand, the $T_{N2}$ transition to the HM phase can be associated with a local minimum in the χ vs. T dependence.

The AC magnetic susceptibility of $NiBr_2$ under pressure is shown in Fig. 2a. $T_{N1}$ exhibits a systematic shift toward higher temperatures with increasing pressure, attaining 99 K at 2.8 GPa. The $T_{N1}$ vs. *p* dependence slope is ~20 K/GPa and shows no sign of saturation, even at the highest pressure. The reported increment in AFM phase with applied pressure is the highest yet observed in any known vdW AFM compounds.

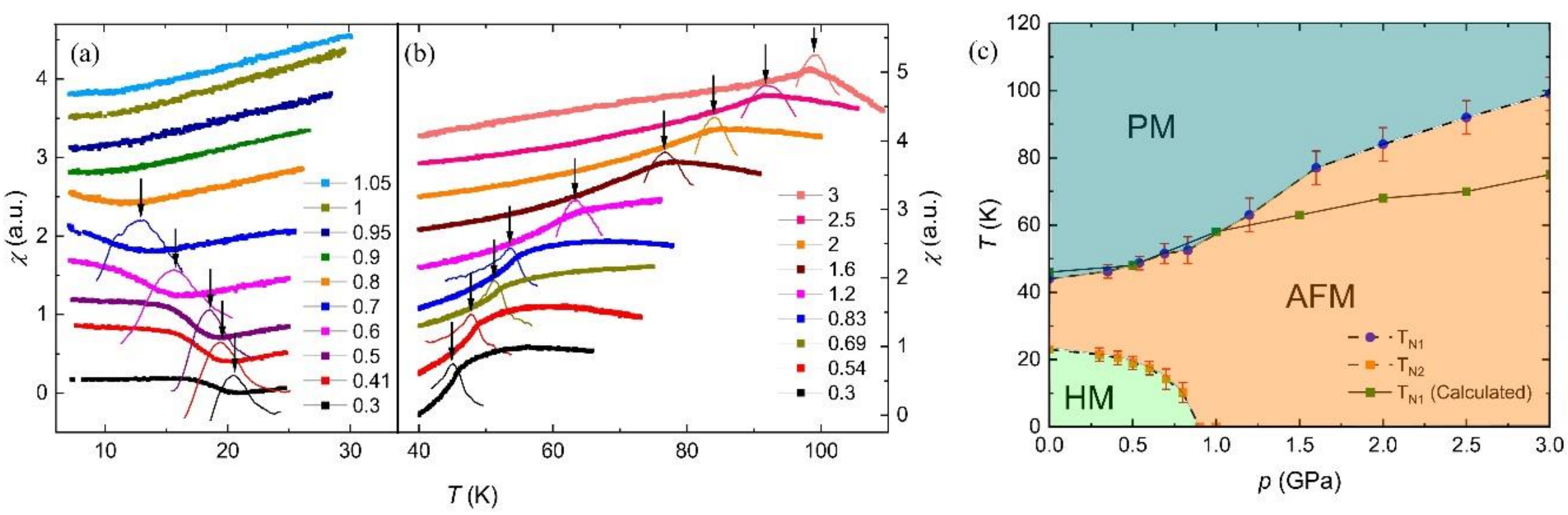

**Fig. 2 | Magnetism of $NiBr_2$ under pressure.** (a,b) Temperature dependence of AC magnetic susceptibility and its temperature derivative (thin lines) at various pressures down to a temperature of 7 K, (a) for the evolution of $T_{N2}$ and (b) for $T_{N1}$. $T_{N1}$ is identified by the maximum in ($d\chi/dT$), and $T_{N2}$ by the maximum ($d^2\chi/dT^2$). The legend is in the GPa scale. An AC magnetic field was applied perpendicular to the *c*-axis with amplitude < 1 mT. (c) *p-T* phase diagram of $NiBr_2$. Symbols (●), (■) and (■) represent the evolution of $T_{N1}$, $T_{N2}$ and $T_{N1}$(calculated), respectively.

On the other hand, $T_{N2}$ rapidly decreases with increasing pressure and vanishes beyond 0.8 GPa. Our measurements provide the first direct determination of the critical pressure for the disappearance of the helimagnetic phase, whereas earlier studies estimated this value to be above 1.06 GPa based on extrapolation from measurements limited to pressure 0.5 GPa and temperature 10 K[24]. The corresponding magnetic *p-T* phase diagram is constructed in Fig. 2c. A distinctive feature is the enhanced rate of the pressure-induced growth of $T_{N1}$ above 0.8 GPa, occurring at the same pressure at which the helical phase vanishes. The decompression process is shown in the *p*-*T* compression-decompression phase diagram in Fig. S2 in Supporting Information (SI), which shows hysteresis for both critical temperatures. As the sample is decompressed to ambient pressure, both values are within the measurement error bars. The hysteresis for $T_{N1}$ and reappearance of $T_{N2}$ may suggest microscopic changes in the $NiBr_2$ crystal structure that influence magnetism. Detailed pressure-dependent structural studies are required to clarify the precise mechanism underlying this behavior.

**Finite-Temperature Magnetism in Layered $NiBr_2$ Using Atomistic Simulations**

We construct an effective isotropic Heisenberg Hamiltonian accounting for both intralayer ($H_{\parallel}$, interlayer ($H_{\perp}$ exchange interactions up to the 3rd in-plane and out-of-plane neighbors. Together with the single-ion-anisotropy ($H_{SIA}$ , the Hamiltonian is given as:

$$H = H_{\parallel} + H_{\perp} + H_{SIA} = \frac{-1}{2}\sum_{l}\sum_{i,j} j_{i,j}\hat{e}_{l,i}\cdot\hat{e}_{l,j} - \sum_{l}\sum_{i,j} j'_{i,j}\,\hat{e}_{l,i}\cdot\hat{e}_{l+1,j} + \sum_{l}\sum_{i} E_{SIA}\left(\hat{e}_{l,i}\right) \quad (1)$$

Here, the local magnetic moment unit vector of the $i^{th}$ magnetic ion in the $l^{th}$ layer is denoted by $\hat{e}_{l,i} = m_{l,i}/\left|m_{l,i}\right|$. The intralayer exchange interactions within the $l^{th}$ layer are characterized by $j'_{i,j}$, with $j'_1, j'_2$ and $j'_3$ denoting the first, second, and third nearest-neighbour interlayer couplings, as shown in Fig. 3b. The single-ion anisotropy energy is expressed as $E_{SIA}\left(\hat{e}_{l,i}\right) = K_U\left(\hat{e}_{l,i}\cdot\hat{e}^{K}\right)^2$where $K_U$ is the first-order uniaxial anisotropy constant, and $\hat{e}^{K}$ is a unit vector defining the axis of the uniaxial anisotropy.

We have calculated relevant interactions using first-principles methods and simulated spin dynamics of this Hamiltonian (see Methods for details), focusing on the formation of HM order. All calculated $j'_{ij}$ are listed in Table S1 in the SI. The calculated exchange parameters match the values obtained from neutron-scattering studies of $NiBr_2$ single crystals[29, 30] within 5% error for the dominant term $j_1$, and 20% error for the sum of all interlayer interactions. This comprehensive approach captures critical magnetic interactions and stochastic effects, enabling accurate modelling of finite-temperature magnetism in layered magnetic systems.

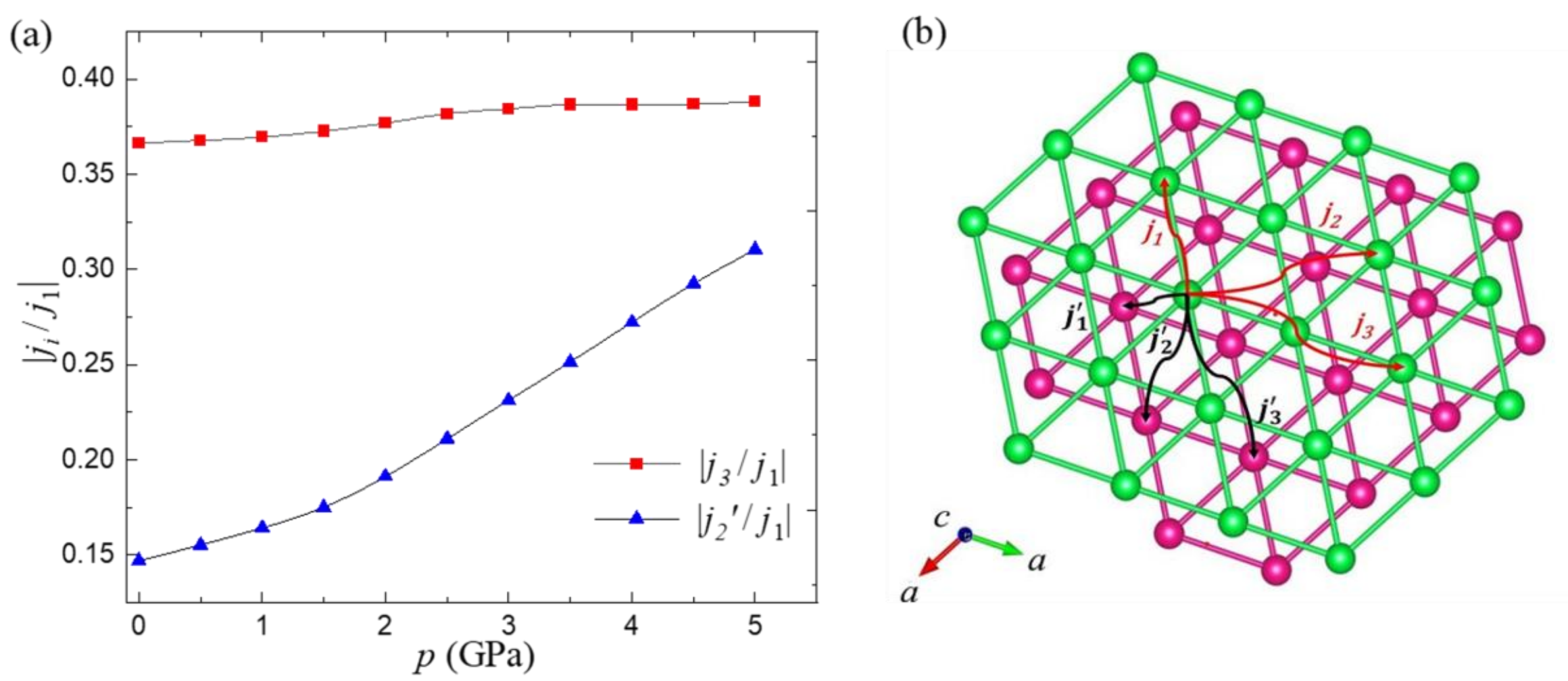

**Fig.3 | Calculated pressure-dependent exchange interactions: a.** Ratios between selected exchange interactions and the dominant interaction $j_1$ as a function of pressure. **b.** The paths for the interlayer exchange interactions $j_1 j_2 j_3$ and intralayer exchange interactions $j'_1, j'_2$ and $j'_3$ are shown over the triangular arrangement of Ni atoms in $NiBr_2$ (with magenta spheres corresponding to Ni atoms in the bottom layer and green spheres to those in the top layer). Local interactions are shown in-plane $a, b\left(|\vec{a}| = \left|\vec{b}\right| = a\right)$ and implied perpendicular *c*.

Within one layer, the mean-field model predicts that at zero temperature, the HM order is preferred over the collinear AFM one if $|j_3| > \frac{j_1}{4}$ [26]. In this case, a transition to collinear order might occur at finite temperature, depending on how much $|j_3|$ differs from $\frac{j_1}{4}$[31]. The mere decrease of $|j_3/j_1|$ could therefore explain the decrease of $T_{N2}$ within the simplest model. Our calculations, however, show that it is not the case for $NiBr_2$; this ratio changes only slightly over the interval from 0 to 5 GPa (Fig. 3a). The evolution of fundamental in-plane interactions is not sufficient to explain the decrease of $T_{N2}$. Therefore, a more complex description of the problem is needed.

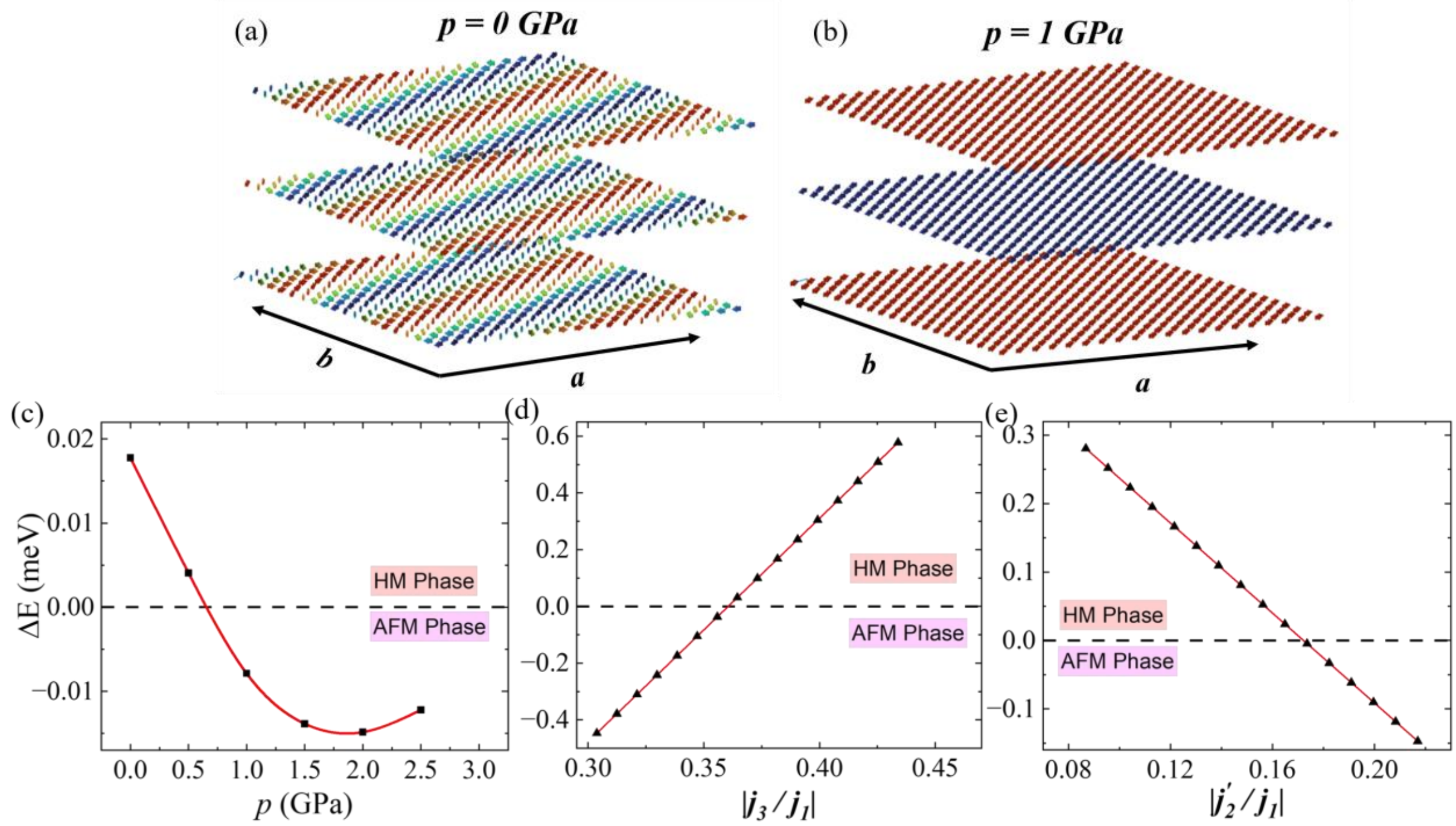

**Fig. 4| Ground State Magnetic Structure and Energy difference (ΔE) between the AFM and HM phase: a.** HM order at ambient pressure with spin spiral ($q \sim 0.25\pi/a$). **b.** AFM order at 1 GPa, with FM alignment within layers and AFM coupling between them. **c.** Energy difference (ΔE) between the AFM and HM phase (positive favors HM) as a function of pressure employing ab initio calculated exchange interactions for each pressure. **d.** Energy difference (ΔE) between the AFM and HM phase obtained for the case when the exchange interaction $j_3$ was varied, while all other parameters were obtained from the ab initio calculation for $p = 0.5$ GPa **e**, the same as **d**, but the exchange interaction $j_2'$ was varied.

We analyze the ground state spin configuration obtained in simulations employing first-principles calculated parameters for different pressures, while the experimentally predicted anisotropy $K_U = 0.15$ meV[29] was used. The predicted spin configuration for ambient pressure exhibits the character of HM order (Fig. 4a). A detailed analysis of spin-spin correlations shows that moments form a spin spiral with $q \sim 0.25\pi/a$. At sufficiently high pressure, the preferred spin configuration becomes FM within layers and AFM between them (layered AFM, Fig. 4b), resulting in zero total magnetization. This behavior is in excellent agreement with the experiment and corroborates that the Hamiltonian (Eq. 1) with the calculated pressure-dependent parameters describes the key physics of the problem.

Since Monte Carlo simulations are not guaranteed to reach the ground state, especially when multiple configurations are close in energy, we also evaluate the energies of the plausible spin configurations (HM, AFM) for all parameter sets obtained for the considered pressures. The difference between the calculated HM and AFM energies as a function of pressure is shown in Fig. 4c. Notably, the transition from the preference for the HM to the AFM order occurs between 0 and 1 GPa, as expected from the findings above.

To explore how specific interactions influence the stability of the magnetic phases under varying conditions, we have studied sets of Hamiltonians where one parameter was varied, while others were held fixed at values calculated by DFT for the selected pressure of 0.5 GPa. We varied the third-nearest-neighbor intralayer exchange ($j_3$) to assess its impact: The energy difference ($\Delta$E) between the AFM and HM magnetic orders for such Hamiltonians at zero temperature is shown in Fig. 4(d). This energy difference scales linearly with the ratio between the selected exchange parameters and the dominant one $j_1$. For parameters calculated for the pressure of 0.5 GPa, changing $j_3$ so that $j_3/j_1 \vee < 0.36$ would lead to the preference of AFM order (and vice versa). The trend thus confirms the expectations of the one-layer mean field model, but the threshold is shifted to a higher value.

Interlayer interactions employ the Ni–Br–Br–Ni superexchange path, where the coupling is suppressed by the need to cross the vdW gap. The reduction in the gap thus has a strong impact on these interactions. Their magnitude will increase with pressure, unless terms originating from different orbitals cancel, which could occur for specific bond angles. The AFM-stacked FM layers are clearly energetically most favorable when this interaction is negative. Therefore, one can expect these interactions to shift the balance between the HM and AFM orders. Among interlayer interactions, $j_2'$ achieves the highest values (Table S1 in the SI). The ratio $|j_2'/j_1|$ increases by more than 150% as pressure is raised to 5 GPa, representing a substantially stronger pressure response than that of the other interlayer couplings (Fig. 3a). Therefore, as the second parameter to be varied, we select $j_2'$, while keeping the other parameters fixed. The resulting energy difference ($\Delta$E) is shown in Fig. 4e. The preferred ground state spin configuration switches from HM to AFM when $|j_2'/j_1| < 0.17$. These results highlight the critical role of interlayer coupling in governing magnetic phase behavior in $NiBr_2$. The slopes allow us to predict the threshold for the simultaneous change of both $j_3, j_2'$ parameters.

The HM order in $NiI_2$ is much more stable than in $NiBr_2$, as the ratio between its critical temperatures ($T_{N2}/T_{N1}$) is ~ 4/5, while in the case of $NiBr_2$ it is ~ 1/2. The HM order is not suppressed even at pressures up to 7 GPa, and $T_{N2}$ evolves oppositely to $NiBr_2$[15]. This difference can be attributed primarily to the different $|j_3/j_1|$ ratio. In the case of $NiBr_2$, this ratio is close to the critical value of 0.25 for the occurrence of HM order[26], while for $NiI_2$ it is claimed to be much higher, 0.81[32]. If we neglect the change of all other parameters between, our simulation results predict that a very high $j_2'$ would be needed to turn it into an AFM state. This comparison shows that, despite chemical similarities, the pressure responses of the magnetic phases in $NiBr_2$ and $NiI_2$ differ due to distinct exchange coupling values.

Néel temperature $T_{N1}$ is predicted to increase significantly with pressure over the studied interval, in agreement with the measurement. Increasing intralayer and interlayer interaction stabilizes magnetic order; this behavior is qualitatively expected. Quantitatively, the predicted increase is slower than the experimentally observed one, but significantly faster than in the case computed previously[25].

The enhanced increase of $T_{N1}$ between ~1 and 2 GPa is not fully captured by first-principles calculations. This may be attributed to multiple contributing factors. Accurately capturing the pressure-induced change in $T_{N1}$ hinges on finding lattice parameters for which interatomic forces are balanced by the applied external pressure. In the case of vdW materials, the key is the use of a vdW functional, which accounts for London dispersion forces, unlike approaches based on pure LDA or GGA. Nevertheless, the development of these methods is still in progress, and errors of several % cannot be excluded in the determination of lattice parameters in layered materials[33]. Furthermore, discrepancies between calculation and experiment for materials with a significant role of vdW interactions would be further enhanced under pressure, as recently demonstrated in a study on ice[34].

Another possible explanation is that the pressure will initiate the formation of stacking faults. If these were to lead to a decrease in $j_3$ faster than $j_1$, the Néel temperature would increase. Multiple other defects could occur, including stacking faults, lattice distortions, etc. More data on the structure would be needed before addressing the problem with first-principles calculations, due to the high number of degrees of freedom.

# CONCLUSION

Our integrated experimental and theoretical investigation demonstrates that hydrostatic pressure serves as a powerful control knob for tuning magnetic phases in $NiBr_2$. Using AC magnetic susceptibility measurements, we observed a continuous increase in the Néel temperature from 44 K at ambient pressure to nearly 100 K at 3 GPa, without signs of saturation. Conversely, the HM phase is rapidly suppressed by pressure, disappearing beyond 0.8 GPa. Atomistic Monte Carlo simulations, supported by first-principles input, have shown that the HM to AFM transition in $NiBr_2$ can be linked to specific parameters of the isotropic Heisenberg Hamiltonian, namely those corresponding to interlayer interactions. The calculations predict a significant increase in the second-nearest interlayer coupling ($j_2'$, which is found to selectively stabilize the collinear AFM state while destabilizing the non-collinear helical order. Our work provides spin models for $NiBr_2$ at different pressures, with $T_{N1}$ prediction errors ranging from a few % at ambient pressure to 30% at high pressures. The HM to AFM transition is predicted in the same pressure range as experimentally observed. This accuracy in spin model construction has been achieved employing the advanced meta-GGA functional r$^2$SCAN together with a nonlocal vdW interaction description. In contrast to $NiI_2$, where the stability of helimagnetic order increases with pressure (within the range studied here), the HM order in $NiBr_2$, is much more fragile. This difference can be ascribed mainly to the different ratio of the intralayer couplings that compete with the interlayer couplings stabilizing the collinear AFM order in these closely related vdW magnets.

## ACKNOWLEDGMENT

This work is a part of the research project GAČR 25-15448S, funded by the Czech Science Foundation, and the project GAUK 128624, funded by the Charles University Grant Agency. The single-crystal growth, characterization, and experiments in steady magnetic fields were carried out in the Materials Growth and Measurement Laboratory (MGML, http://mgml.eu), supported within the program of Czech Research Infrastructures (project no. LM2023065). K.C., K.P., and S.R. acknowledge support from the project Quantum Materials for Applications in Sustainable Technologies (QM4ST), funded as project no. CZ.02.01.01/00/22_008/0004572 by P JAK under the call Excellent Research. This work was also supported by the Ministry of Education, Youth and Sports of the Czech Republic through the e-INFRA CZ (ID: 90140).

## METHODS

### Single crystal growth and characterization

The single crystals of $NiBr_2$ have been grown from pure elements (Ni 99.95 %, $Br_2$ 99.5%) using the chemical vapor transport method, using a similar procedure we recently invented to prepare other Br-based vdW materials[35]. This approach prevented contamination of the final product by residuals from precursors readily used to produce $Br_2$, e.g., $TeBr_4$, in the reaction and transporting tube. The sealed quartz tube was inserted into a gradient furnace where a 600/500°C thermal gradient was maintained for two weeks to transport all Ni metal from the hot part. The single crystals of transparent yellow-brown color, typically centimeter-sized, have been obtained. The desired 1:2 composition was confirmed by EDX analysis. The crystallinity and orientation of the single crystals were confirmed by Laue method, which showed sharp reflections. The hexagonal *c*-axis is perpendicular to the plane of plate-like crystals.

### Pressure Experiment

The effect of hydrostatic pressure on the magnetism in $NiBr_2$ was detected using the AC magnetic susceptibility method with a custom-built setup comprising miniature coils, as described in more detail in[36]. These coils have a diameter of approximately 1.5 mm and a length of around 6 mm. The primary coil, which generates the AC magnetic field, consists of about 200 turns. Meanwhile, the secondary coil, used to detect the AC magnetic susceptibility, features two sections with 50 turns each, wound in opposite directions to compensate for the signal of the sample's surroundings. The coils were interfaced with a Stanford Research Systems SR 830 lock-in amplifier for AC signal processing to measure the real ($\chi'$) and imaginary ($\chi''$) components. The coil assembly is housed within a double-layered CuBe/NiCrAl piston-cylinder pressure cell[37], capable of generating pressures up to approximately 3.5 GPa at room temperature. Daphne oil 7474[38, 39] was used as the pressure-transmitting medium, and a manganin manometer was employed to determine the pressure at room temperature. After cooling to the base temperature, the pressure was corrected for the pressure change caused by the thermal expansion of the cell and the Daphne oil 7474. The AC magnetic susceptibility measurements at low temperatures down to 3 K were performed using a closed-cycle refrigerator (Janis Research Company, LLC/Sumitomo Heavy Industries, Ltd.). The experimental AC magnetic susceptibility data are shown only down to 7 K, because the amplified artefact of the superconducting transition of the lead (materials of solder) appears below this temperature. The reference ambient-pressure specific heat data were collected over a 2-300 K temperature range in a PPMS-14 system (Quantum Design, Inc.), and the corresponding AC magnetic susceptibility was measured in an MPMS 7T.

## COMPUTATIONAL METHODS

### Atomistic Simulations of Finite-Temperature Magnetism in Layered Systems

The magnetic behavior of $NiBr_2$ at varying temperatures was explored through atomistic simulations using the Uppsala Atomistic Spin Dynamics (UppASD) framework[40-42]. The system studied comprised 20×20×10-unit cells, and simulations included 90,000 Monte Carlo (MC) steps per temperature point to achieve equilibrium, followed by 90,000 additional spin-dynamics steps to calculate average magnetization. Each temperature configuration was run three times and averaged to ensure statistical accuracy. The simulations relied on solving the Landau-Lifshitz-Gilbert (LLG) equation with a fine time step of 0.1 femtoseconds. Following the fluctuation-dissipation theorem, they incorporated random thermal fluctuations via Langevin dynamics to mimic temperature-driven effects[40]. The Néel temperature $T_{N1}$ was determined from the maximum of the computed heat capacity curve, shown in Fig. S1 in the SI.

#### *Heisenberg exchange interactions by mapping the DFT*

Evaluation of exchange interactions is generally a complex problem, and methods that can handle multiple conditions are still being developed[43]. We estimate isotropic Heisenberg exchange interactions by mapping the DFT energies of supercells with broken-symmetry magnetic configurations to the Heisenberg model. We assume that the magnetic energy levels of a magnetic system μ can be described by a well-defined set of spin-exchange parameters[44]

$$E_M^\mu = \frac{-1}{2}\sum_{j\neq i} J_{i,j} S_i^\mu S_j^\mu \qquad (2)$$

where, $E_M^\mu$ is the magnetic contribution to the total DFT energy $E_0^\mu$, $j_{i,j}$ are isotropic exchange interactions between spins $S_i^\mu$ and $S_j^\mu$ located at sites $i$ and $j$, respectively ($|S_i^\mu| > 1$. In this approximation, the isotropic Heisenberg exchange interactions $J_{i,j}$ can be estimated by mapping the set of $N_\mu$ Eq. 2 to the DFT calculated energies $E_M^\mu$ of a set of $N_\mu$ magnetic configurations with broken symmetry μ[36]. The set of magnetic phases with broken symmetry μ consists of the FM cell and a set of supercells with various AFM and uncompensated antiferromagnetic (or ferrimagnetic) (uAFM) alignments of spins. Electronic structure calculations of the set of $N_\mu$magnetic configurations are performed for a set of volumes, and the Heisenberg interactions $J_{i,j}$ are calculated at each volume. In this work, we considered up to NN=8 nearest neighbor shells about each Ni atom, and so, the Heisenberg sum in Eq.1 can be written as

$$\sum_{j\neq i} J_{ij} S_i^\mu S_j^\mu = \sum_j^{NN} J_{0j} S_0 \sum_k^{Z_j^\mu} S_k = \sum_j^{NN} J_{0j} S_{0j}^\mu \qquad (3)$$

where, $S^{\mu}_{0j} = S_0 \sum_k^{Z_j^{\mu}} S_k$ with summation over all $Z_j^{\mu}$ spins in the shell $j$ for the magnetic configuration μ. Within the six calculated $J_{i,j}$ values for the layered $NiBr_2$, three will correspond to the intralayer ($j_{i,j}$, and three to the interlayer ($j'_{i,j}$ exchange parameters. All calculated parameters are compiled in Table S1 of the Supporting Information; two of them are shown in (Fig. 3b).

All magnetic configurations were calculated with the VASP software (VASP version 6.4.3), which is a plane-wave basis set implementation[45, 46] of the Density Functional Theory (DFT) with the Projector Augmented Wave (PAW) approximation[47]. Calculations were performed using a combination of the $r^2$SCAN meta-generalized gradient approximation[48] to the exchange-correlation (X-C) part of the energy functional and nonlocal rVV10 van der Waals (vdW) functional[49-51]. This method has been shown to perform significantly better than many other combinations of X-C and vdW functionals for hexagonal layered solids (on average), as studied on a set of 10 materials [33]. In addition, to account for strong electron0 0 2 0 correlation effects, we used the simplified rotationally invariant approach of DFT+U[52] with $U$ = 2.5 eV. The critical quantity$j_3/j_1$ changes by less than 3% for U ranging from 2.5 eV to 3.9 eV, while for $j_2'/j_1$ the variation is less than 16%. We used the PAW PBE potential version 5.4 with 16 and 17 valence electrons for Ni and Br, respectively. The size of the plane-wave basis set was determined by an energy cut-off of 644 eV, which is 75% larger than the default one. The reciprocal space was sampled with a uniform mesh of k-points[53] with a separation between k-points of 0.1 $Å^{-1}$. For the total energy summation, we used the Methfessel-Paxton method[54]The smearing width was 0.05 eV, and the electronic convergence was set to $10^{-7}$ eV. The structural optimization of the AFM configuration (the 1×1×2 cell with antiferromagnetically alternating ferromagnetic layers of Ni). The geometry optimization was performed until the norms of all of the Hellman-Feynman forces[55] became less than $10^{-3}$ eV/Å. Equilibrium structure parameters were obtained by fitting the energy vs volume data E = E(V) to the Rose-Vinet equation of state (EOS)[56].

## References:

1. A.K. Geim and K.S. Novoselov. The rise of graphene. *Nature Materials* 6 (2007) 183-191, https://doi.org/10.1038/nmat1849
2. Q.H. Wang, A. Bedoya-Pinto, M. Blei, A.H. Dismukes, A. Hamo, S. Jenkins, M. Koperski, Y. Liu, Q.-C. Sun, E.J. Telford, H.H. Kim, M. Augustin, U. Vool, J.-X. Yin, L.H. Li, A. Falin, C.R. Dean, F. Casanova, R.F.L. Evans, M. Chshiev, A. Mishchenko, C. Petrovic, R. He, L. Zhao, A.W. Tsen, B.D. Gerardot, M. Brotons-Gisbert, Z. Guguchia, X. Roy, S. Tongay, Z. Wang, M.Z. Hasan, J. Wrachtrup, A. Yacoby, A. Fert, S. Parkin, K.S. Novoselov, P. Dai, L. Balicas, and E.J.G. Santos. The Magnetic Genome of Two-Dimensional van der Waals Materials. *ACS Nano* 16 (2022) 6960-7079, https://doi.org/10.1021/acsnano.1c09150
3. C. Huang, J. Zhou, H. Sun, F. Wu, Y. Hou, and E. Kan. Toward Room-Temperature Electrical Control of Magnetic Order in Multiferroic van der Waals Materials. *Nano Letters* 22 (2022) 5191-5197, https://doi.org/10.1021/acs.nanolett.2c00930
4. Q. Song, C.A. Occhialini, E. Ergeçen, B. Ilyas, D. Amoroso, P. Barone, J. Kapeghian, K. Watanabe, T. Taniguchi, A.S. Botana, S. Picozzi, N. Gedik, and R. Comin. Evidence for a single-layer van der Waals multiferroic. *Nature* 602 (2022) 601-605, https://doi.org/10.1038/s41586-021-04337-x
5. S. Yuan, X. Luo, H.L. Chan, C. Xiao, Y. Dai, M. Xie, and J. Hao. Room-temperature ferroelectricity in MoTe2 down to the atomic monolayer limit. *Nature Communications* 10 (2019) 1775, https://doi.org/10.1038/s41467-019-09669-x
6. N. Higashitarumizu, H. Kawamoto, C.-J. Lee, B.-H. Lin, F.-H. Chu, I. Yonemori, T. Nishimura, K. Wakabayashi, W.-H. Chang, and K. Nagashio. Purely in-plane ferroelectricity in monolayer SnS at room temperature. *Nature Communications* 11 (2020) 2428, https://doi.org/10.1038/s41467-020-16291-9
7. X. Wang, Z. Shang, C. Zhang, J. Kang, T. Liu, X. Wang, S. Chen, H. Liu, W. Tang, Y.-J. Zeng, J. Guo, Z. Cheng, L. Liu, D. Pan, S. Tong, B. Wu, Y. Xie, G. Wang, J. Deng, T. Zhai, H.-X. Deng, J. Hong, and J. Zhao. Electrical and magnetic anisotropies in van der Waals multiferroic

$CuCrP_2S_6$. *Nature Communications* 14 (2023) 840, https://doi.org/10.1038/s41467-023-36512-1

8. L. Cao, X. Deng, G. Zhou, S.-J. Liang, C.V. Nguyen, L.K. Ang, and Y.S. Ang. Multiferroic van der Waals heterostructure $FeCl_2/Sc_2CO_2$: Nonvolatile electrically switchable electronic and spintronic properties. *Physical Review B* 105 (2022) 165302, https://doi.org/10.1103/PhysRevB.105.165302
9. M. Bonilla, S. Kolekar, Y. Ma, H.C. Diaz, V. Kalappattil, R. Das, T. Eggers, H.R. Gutierrez, M.-H. Phan, and M. Batzill. Strong room-temperature ferromagnetism in $VSe_2$ monolayers on van der Waals substrates. *Nature Nanotechnology* 13 (2018) 289-293, https://doi.org/10.1038/s41565-018-0063-9
10. Y. Deng, Y. Yu, Y. Song, J. Zhang, N.Z. Wang, Z. Sun, Y. Yi, Y.Z. Wu, S. Wu, J. Zhu, J. Wang, X.H. Chen, and Y. Zhang. Gate-tunable room-temperature ferromagnetism in two-dimensional $Fe_3GeTe_2$. *Nature* 563 (2018) 94-99, https://doi.org/10.1038/s41586-018-0626-9
11. C. Gong, L. Li, Z. Li, H. Ji, A. Stern, Y. Xia, T. Cao, W. Bao, C. Wang, Y. Wang, Z.Q. Qiu, R.J. Cava, S.G. Louie, J. Xia, and X. Zhang. Discovery of intrinsic ferromagnetism in two-dimensional van der Waals crystals. *Nature* 546 (2017) 265-269, https://doi.org/10.1038/nature22060
12. P.W. Anderson. Antiferromagnetism. Theory of Superexchange Interaction. *Physical Review* 79 (1950) 350-356, https://doi.org/10.1103/PhysRev.79.350
13. D.P. Kozlenko, O.N. Lis, N.T. Dang, M. Coak, J.G. Park, E.V. Lukin, S.E. Kichanov, N.O. Golosova, I.Y. Zel, and B.N. Savenko. High-pressure effects on structural, magnetic, and vibrational properties of van der Waals antiferromagnet $MnPS_3$. *Physical Review Materials* 8 (2024) 024402, https://doi.org/10.1103/PhysRevMaterials.8.024402
14. M.J. Coak, D.M. Jarvis, H. Hamidov, A.R. Wildes, J.A.M. Paddison, C. Liu, C.R.S. Haines, N.T. Dang, S.E. Kichanov, B.N. Savenko, S. Lee, M. Kratochvílová, S. Klotz, T.C. Hansen, D.P. Kozlenko, J.-G. Park, and S.S. Saxena. Emergent Magnetic Phases in Pressure-Tuned van der Waals Antiferromagnet $FePS_3$. *Physical Review X* 11 (2021) 011024, https://doi.org/10.1103/PhysRevX.11.011024
15. Q. Liu, W. Su, Y. Gu, X. Zhang, X. Xia, L. Wang, K. Xiao, N. Zhang, X. Cui, M. Huang, C. Wei, X. Zou, B. Xi, J.-W. Mei, and J.-F. Dai. Surprising pressure-induced magnetic transformations from helimagnetic order to antiferromagnetic state in $NiI_2$. *Nature Communications* 16 (2025) 4221, https://doi.org/10.1038/s41467-025-59561-0
16. D. Billerey, C. Terrier, N. Ciret, and J. Kleinclauss. Neutron diffraction study and specific heat of antiferromagnetic $NiI_2$. *Physics Letters A* 61 (1977) 138-140, https://doi.org/10.1016/0375-9601(77)90863-5
17. T. Kurumaji, S. Seki, S. Ishiwata, H. Murakawa, Y. Kaneko, and Y. Tokura. Magnetoelectric responses induced by domain rearrangement and spin structural change in triangular-lattice helimagnets $Ni_2$ and $CoI_2$. *Physical Review B* 87 (2013) 014429, https://doi.org/10.1103/PhysRevB.87.014429
18. I. Tsubokawa. The Magnetic Properties of $NiBr_2$. *Journal of the Physical Society of Japan* 15 (1960) 2109-2109, https://doi.org/10.1143/JPSJ.15.2109
19. Q. Cui, L. Wang, Y. Zhu, J. Liang, and H. Yang. Magnetic anisotropy, exchange coupling and Dzyaloshinskii—Moriya interaction of two-dimensional magnets. *Frontiers of Physics* 18 (2022) 13602, https://doi.org/10.1007/s11467-022-1217-7
20. S. Babu, K. Prokeš, Y.K. Huang, F. Radu, and S.K. Mishra. Magnetic-field-induced incommensurate to collinear spin order transition in $NiBr_2$. *Journal of Applied Physics* 125 (2019) 093902, https://doi.org/10.1063/1.5066625
21. Y. Tokunaga, D. Okuyama, T. Kurumaji, T. Arima, H. Nakao, Y. Murakami, Y. Taguchi, and Y. Tokura. Multiferroicity in $NiBr_2$ with long-wavelength cycloidal spin structure on a triangular lattice. *Physical Review B* 84 (2011) 060406, https://doi.org/10.1103/PhysRevB.84.060406
22. A. Adam, D. Billerey, C. Terrier, R. Mainard, L.P. Regnault, J. Rossat-Mignod, and P. Mériel. Neutron diffraction study of the commensurate and incommensurate magnetic structures of $NiBr_2$. *Solid State Communications* 35 (1980) 1-5, https://doi.org/10.1016/0038-1098(80)90757-7

23. P. Day and C. Vettier. Pressure dependence of the incommensurate magnetic phase of $NiBr_2$ and $Ni_{0.92}Zn_{0.08}Br_2$. *Journal of Physics C: Solid State Physics* 14 (1981) L195, https://doi.org/10.1088/0022-3719/14/8/006
24. A. Adam, D. Billerey, C. Terrier, H. Bartholin, L.P. Regnault, and J. Rossat-Mignod. Hydrostatic pressure effect on the commensurate—incommensurate phase transition of $NiBr_2$. *Physics Letters A* 84 (1981) 24-27, https://doi.org/10.1016/0375-9601(81)90008-6
25. S. Bag, J. Kapeghian, O. Erten, and A.S. Botana. Tuning the electronic and magnetic properties of $NiBr_2$ via pressure. *Physical Review B* 110 (2024) 085161, https://doi.org/10.1103/PhysRevB.110.085161
26. E. Rastelli, A. Tassi, and L. Reatto. Non-simple magnetic order for simple Hamiltonians. *Physica B+C* 97 (1979) 1-24, https://doi.org/10.1016/0378-4363(79)90002-0
27. X. Li, C. Xu, B. Liu, X. Li, L. Bellaiche, and H. Xiang. Realistic Spin Model for Multiferroic $NiI_2$. *Physical Review Letters* 131 (2023) 036701, https://doi.org/10.1103/PhysRevLett.131.036701
28. B.K. Rai, S. Gao, M. Frontzek, Y. Liu, A.D. Christianson, and A.F. May. Magnetic properties of Fe-substituted $NiBr_2$ single crystals. *Journal of Magnetism and Magnetic Materials* 557 (2022) 169452, https://doi.org/10.1016/j.jmmm.2022.169452
29. J.R.-M. L.P. Régnault, A. Adam, D. Billerey, C. Terrier. Inelastic neutron scattering investigation of the magnetic excitations in the helimagnetic state of $NiBr_2$. *Journal de Physique* 43 ( 1982) 1283-1290, https://doi.org/10.1051/jphys:019820043080128300
30. P. Day, M.W. Moore, T.E. Wood, D.M. Paul, K.R.A. Ziebeck, L.P. Regnault, and J. Rossat-Mignod. Inelastic neutron scattering study of magnetic excitations in the helimagnetic and antiferromagnetic phases of NiBr2. *Solid State Communications* 51 (1984) 627-630, https://doi.org/https://doi.org/10.1016/0038-1098(84)91075-5
31. J. Villain. Helimagnetic-ferromagnetic transition. *Physica B+C* 86-88 (1977) 631-633, https://doi.org/https://doi.org/10.1016/0378-4363(77)90626-X
32. J. Kapeghian, D. Amoroso, C.A. Occhialini, L.G.P. Martins, Q. Song, J.S. Smith, J.J. Sanchez, J. Kong, R. Comin, P. Barone, B. Dupé, M.J. Verstraete, and A.S. Botana. Effects of pressure on the electronic and magnetic properties of bulk $NiI_2$. *Physical Review B* 109 (2024) 014403, https://doi.org/10.1103/PhysRevB.109.014403
33. F. Tran, L. Kalantari, B. Traoré, X. Rocquefelte, and P. Blaha. Nonlocal van der Waals functionals for solids: Choosing an appropriate one. *Physical Review Materials* 3 (2019) 063602, https://doi.org/10.1103/PhysRevMaterials.3.063602
34. B. Santra, J. Klimeš, A. Tkatchenko, D. Alfè, B. Slater, A. Michaelides, R. Car, and M. Scheffler. On the accuracy of van der Waals inclusive density-functional theory exchange-correlation functionals for ice at ambient and high pressures. *The Journal of Chemical Physics* 139 (2013) 154702, https://doi.org/10.1063/1.4824481
35. D. Hovančík, M. Kratochvílová, T. Haidamak, P. Doležal, K. Carva, A. Bendová, J. Prokleška, P. Proschek, M. Míšek, D.I. Gorbunov, J. Kotek, V. Sechovský, and J. Pospíšil. Robust intralayer antiferromagnetism and tricriticality in the van der Waals compound $VBr_3$. *Physical Review B* 108 (2023) 104416, https://doi.org/10.1103/PhysRevB.108.104416
36. J. Valenta, M. Kratochvílová, M. Míšek, K. Carva, J. Kaštil, P. Doležal, P. Opletal, P. Čermák, P. Proschek, K. Uhlířová, J. Prchal, M.J. Coak, S. Son, J.G. Park, and V. Sechovský. Pressure-induced large increase of Curie temperature of the van der Waals ferromagnet $VI_3$. *Physical Review B* 103 (2021) 054424, https://doi.org/10.1103/PhysRevB.103.054424
37. N. Fujiwara, T. Matsumoto, K. Nakazawab, A. Hisada, and Y. Uwatoko. Fabrication and efficiency evaluation of a hybrid NiCrAl pressure cell up to 4GPa. *Review of Scientific Instruments* 78 (2007) 073905, https://doi.org/10.1063/1.2757129
38. D. Staško, J. Prchal, M. Klicpera, S. Aoki, and K. Murata. Pressure media for high pressure experiments, Daphne Oil 7000 series. *High Pressure Research* 40 (2020) 525-536, https://doi.org/10.1080/08957959.2020.1825706
39. K. Murata, K. Yokogawa, H. Yoshino, S. Klotz, P. Munsch, A. Irizawa, M. Nishiyama, K. Iizuka, T. Nanba, T. Okada, Y. Shiraga, and S. Aoyama. Pressure transmitting medium Daphne 7474 solidifying at 3.7 GPa at room temperature. *Review of Scientific Instruments* 79 (2008) 085101, https://doi.org/10.1063/1.2964117
40. O. Eriksson, A. Bergman, L. Bergqvist, and J. Hellsvik *Atomistic Spin Dynamics: Foundations and Applications*. 2017. https://doi.org/10.1093/oso/9780198788669.001.0001
41. B. Skubic, J. Hellsvik, L. Nordström, and O. Eriksson. A method for atomistic spin dynamics simulations: implementation and examples. *Journal of Physics: Condensed Matter* 20 (2008) 315203, https://doi.org/10.1088/0953-8984/20/31/315203
42. *Uppsala Atomistic Spin Dynamics (UppASD) code. Available under GNU General Public Licence* github. http://physics.uu.se/uppsd/ and http://github.com/UppASD/UppASD/. (accessed).
43. A. Szilva, Y. Kvashnin, E.A. Stepanov, L. Nordström, O. Eriksson, A.I. Lichtenstein, and M.I. Katsnelson. Quantitative theory of magnetic interactions in solids. *Reviews of Modern Physics* 95 (2023) 035004, https://doi.org/10.1103/RevModPhys.95.035004
44. H. Xiang, C. Lee, H.-J. Koo, X. Gong, and M.-H. Whangbo. Magnetic properties and energy-mapping analysis. *Dalton Transactions* 42 (2013) 823-853, https://doi.org/10.1039/C2DT31662E
45. G. Kresse and D. Joubert. From ultrasoft pseudopotentials to the projector augmented-wave method. *Physical Review B* 59 (1999) 1758-1775, https://doi.org/10.1103/PhysRevB.59.1758
46. G. Kresse and J. Furthmüller. Efficiency of ab-initio total energy calculations for metals and semiconductors using a plane-wave basis set. *Computational Materials Science* 6 (1996) 15-50, https://doi.org/10.1016/0927-0256(96)00008-0
47. P.E. Blöchl. Projector augmented-wave method. *Physical Review B* 50 (1994) 17953-17979, https://doi.org/10.1103/PhysRevB.50.17953
48. J.W. Furness, A.D. Kaplan, J. Ning, J.P. Perdew, and J. Sun. Accurate and Numerically Efficient r2SCAN Meta-Generalized Gradient Approximation. *The Journal of Physical Chemistry Letters* 11 (2020) 8208-8215, https://doi.org/10.1021/acs.jpclett.0c02405
49. J. Klimeš, D.R. Bowler, and A. Michaelides. Van der Waals density functionals applied to solids. *Physical Review B* 83 (2011) 195131, https://doi.org/10.1103/PhysRevB.83.195131
50. J. Klimeš, D.R. Bowler, and A. Michaelides. Chemical accuracy for the van der Waals density functional. *Journal of Physics: Condensed Matter* 22 (2010) 022201, https://doi.org/10.1088/0953-8984/22/2/022201
51. J. Ning, M. Kothakonda, J.W. Furness, A.D. Kaplan, S. Ehlert, J.G. Brandenburg, J.P. Perdew, and J. Sun. Workhorse minimally empirical dispersion-corrected density functional with tests for weakly bound systems: $r^2$SCAN + rVV10. *Physical Review B* 106 (2022) 075422, https://doi.org/10.1103/PhysRevB.106.075422
52. S.L. Dudarev, G.A. Botton, S.Y. Savrasov, C.J. Humphreys, and A.P. Sutton. Electron-energy-loss spectra and the structural stability of nickel oxide: An LSDA+U study. *Physical Review B* 57 (1998) 1505-1509, https://doi.org/10.1103/PhysRevB.57.1505
53. H.J. Monkhorst and J.D. Pack. Special points for Brillouin-zone integrations. *Physical Review B* 13 (1976) 5188-5192, https://doi.org/10.1103/PhysRevB.13.5188
54. M. Methfessel and A.T. Paxton. High-precision sampling for Brillouin-zone integration in metals. *Physical Review B* 40 (1989) 3616-3621, https://doi.org/10.1103/PhysRevB.40.3616
55. R.P. Feynman. Forces in Molecules. *Physical Review* 56 (1939) 340-343, https://doi.org/10.1103/PhysRev.56.340
56. P. Vinet, J.R. Smith, J. Ferrante, and J.H. Rose. Temperature effects on the universal equation of state of solids. *Physical Review B* 35 (1987) 1945-1953, https://doi.org/10.1103/PhysRevB.35.1945

**Additional information**

The online version contains supporting material.

**Data availability**

The online version contains all data as part of the supporting material.

**Author Contribution**

P. A. Q. and J. Pr. performed the pressure experiments and analyzed the data; K. K. P., S. R., S. A. and K. C. performed the theoretical calculation; J. Po. has grown the single crystals; P. A. Q., K. C. V. S. and J. Po. wrote the manuscript, all coauthors revised the manuscript before the submission; J. Po. has coordinated the work.

**Competing interests**

The authors declare no competing interests.

Correspondence and requests for materials should be addressed to Jiří Pospíšil and Karel Carva